\documentclass[10pt,conference]{IEEEtran}

\usepackage{epsfig}
\usepackage{graphicx}
\usepackage{amsmath, amssymb}

\newtheorem{lemma}{Lemma}

\newtheorem{example}{Example}
\newtheorem{corollary}{Corollary}
\newtheorem{proposition}{Proposition}

\begin{document}

\title{Group Secret Key Generation Algorithms}

\author{
\authorblockN{Chunxuan Ye and Alex Reznik}
\authorblockA{InterDigital Communications Corporation \\
King of Prussia, PA 19406\\
Email: \{Chunxuan.Ye, Alex.Reznik\}@interdigital.com}
 }
%

\maketitle

\begin{abstract}
We consider a pair-wise independent network where every pair of
terminals in the network observes a common pair-wise source that is
independent of all the sources accessible to the other pairs. We
propose a method for secret key agreement in such a network that is
based on well-established point-to-point techniques and repeated
application of the one-time pad. Three specific problems are
investigated. 1) Each terminal's observations are correlated only
with the observations of a central terminal. All these terminals
wish to generate a common secret key. 2) In a pair-wise independent
network, two designated terminals wish to generate a secret key with
the help of other terminals. 3) All the terminals in a pair-wise
independent network wish to generate a common secret key. A separate
protocol for each of these problems is proposed. Furthermore, we
show that the protocols for the first two problems are optimal and
the protocol for the third problem is efficient, in terms of the
resulting secret key rates.
\end{abstract}

\section{Introduction}

The problem of secret key generation by two terminals, based on
their respective observations of a common source followed by public
transmissions between them, was first studied by Maurer
\cite{Mau93}, and Ahlswede and Csisz\'{a}r \cite{AhlCsi93}. Various
extensions of this problem have been investigated since then (see,
e.g., \cite{CsiNar00}, \cite{MauWol03}, \cite{YeNar04},
\cite{YeNar05}, \cite{YeRezSha06}).

Csisz\'{a}r and Narayan \cite{CsiNar04} generalize the secret key
generation problem to multiple terminals. They consider a model with
an arbitrary number of terminals, each with distinct observations of
a common source. A group of terminals wish to generate a secret key
with the help of other terminals. In generating such a key, these
terminals are allowed to communicate with each other through a
noiseless public channel.

In this paper, we consider a pair-wise independent network where
every pair of terminals in the network observes a common source that
is independent of all the sources accessible to the other pairs.
This model, as a special case of the model in \cite{CsiNar04}, is
motivated by wireless communications \cite{YeRezSha06},
\cite{YeRezSteSha07}. In a wireless communication environment, each
pair of wireless terminals typically possesses some means of
estimating their mutual channel. The resulting estimates are highly
statistically similar, provided that the terminals communicate on
the same carrier frequency. Moreover, any third terminal's
observations are essentially uncorrelated with the observations of
the first two terminals, provided that the third terminal is located
at least half a wavelength away from those two.

The main contribution of this paper is the following. We propose a
method for secret key agreement in the pair-wise independent network
that is based on well-established point-to-point techniques
\cite{YeNar05}, \cite{YeRezSha06} and repeated application of the
one-time pad. Specifically, we propose protocols for three cases of
the pair-wise independent model and prove that the secret key rates
achieved by our protocols are optimal in the first two cases.
Therefore, the capacity problem in such situations is now solved.
Furthermore, the efficiency of our protocol for the last case is
shown through examples.  The innate connections between the
pair-wise independent network and graphs can be observed through
these protocols.

\section{Preliminaries}
Suppose $m\geq 2$ terminals respectively observe $n$ independent and
identically distributed repetitions of the random variables $(X_1,
X_2,...,X_m)$, denoted by $(X_1^n, X_2^n, \cdots , X_m^n)$ with
$X_i^n=(X_{i,1}, \cdots , X_{i,n})$. A group $A\subseteq \{1,\cdots
,m\}$ of terminals wish to generate a common secret key, with the
help of the remaining terminals. To do so, these $m$ terminals can
communicate with each other through a noiseless public channel. The
generated group secret key $K$ should be nearly statistically
independent of the public transmissions. The entropy rate of the
secret key, viz., $H(K)/n$, is called a secret key rate. The largest
achievable secret key rate is called the secret key capacity,
denoted by $C_{SK}(A)$. It is shown in \cite{CsiNar04} that
\[
C_{SK}(A)= H(X_1, ...,X_m)-\min_{(R_1,...,R_m)\in {\cal
R}(A)}\sum_{i=1}^{m}R_i, 
\]
where
\begin{eqnarray*}
{\cal R}(A)&=&\{(R_1, \cdots , R_m): \sum_{i\in B} R_i\geq H(X_{B}|X_{B^c}), \\
           && B\subset \{1, \cdots, m\}, A\not\subset B \},
\end{eqnarray*}
with $X_{B}=\{X_j:\ j\in B\}$ and $B^c=\{1,\cdots ,m\}\backslash B$.

Let $(B_1,\cdots, B_k)$ be a $k$-partition of $\{1,\cdots, m\}$,
such that each element $B_l$, $1\leq l\leq k$, intersects with the
set $A\subseteq \{1,\cdots ,m\}$. Denote by ${\cal B}_k(A)$ the set
of all such $k$-partitions. Then an upper bound on the secret key
capacity is \cite{CsiNar04}
\begin{equation}
C_{SK}(A)\leq \min_{2\leq k\leq |A|}\frac{1}{k-1}I_k(A),
\label{upperbound}
\end{equation}
where
\[
I_k(A)=\min_{(B_1,\cdots, B_k)\in {\cal
B}_k(A)}\sum_{l=1}^{k}H(X_{B_l})-H(X_1,\cdots, X_m).
\]

\section{A pair-wise independent network}
In this paper, we focus on a pair-wise independent network, which is
a special case of the network described in Section II. Suppose that
the observation $X_i$ by terminal $i$ has $m-1$ components
$(Y_{i,1},\cdots ,Y_{i,i-1}, Y_{i,i+1},\cdots , Y_{i,m})$. Each
component $Y_{i,j}$ denotes the observation of the source that is
accessible only to terminals $i$ and $j$. Furthermore, it is assumed
that
\begin{equation}
I(Y_{i,j}, Y_{j,i};\{Y_{k,l}: (k,l)\neq (i,j), (j,i)\})=0.
\label{independent}
\end{equation}
This implies that each source accessible to a pair of terminals is
independent of all other sources--hence, the network is called
pair-wise independent.

If a group of terminals in the pair-wise independent network
generate a common secret key, then an upper bound on the secret key
capacity is given in the following lemma.

\begin{lemma}
In the pair-wise independent network,
\begin{equation}
C_{SK}(A)\leq \min_{2\leq k\leq |A|}\frac{1}{k-1} I_k'(A),
\label{upperbound2}
\end{equation}
where
\[
I_k'(A)= \min_{(B_1,\cdots, B_k)\in {\cal B}_k(A)}\sum_{i,j: i\in
B_{l}; \atop j\in B_{r}; l<r }I(Y_{i,j}; Y_{j,i}).
\]
\end{lemma}

\begin{proof}
Let $(B_1,\cdots ,B_k)$ be an arbitrary $k$-partition belonging to
${\cal B}_k(A)$. It follows from the independence condition
(\ref{independent}) that
\[
H(X_1,\cdots ,X_m)=\sum_{1\leq i<j\leq m}H(Y_{i,j}, Y_{j,i}),
\]
and for $1\leq l\leq k$,
\[
H(X_{B_l})=\sum_{i,j: i<j; i,j\in B_l}H(Y_{i,j}, Y_{j,i})+\sum_{i,j:
i\in B_l;j\not\in B_l}H(Y_{i,j}).
\]
Then
\begin{eqnarray*}
&&\sum_{l=1}^{k}H(X_{B_l})-H(X_1,\cdots, X_m) \\
&=&\sum_{i,j: i\in B_{l}; \atop j\in B_{r};
l<r}\left[H(Y_{i,j})+H(Y_{j,i})-H(Y_{i,j},Y_{j,i})\right]\\
&=&\sum_{i,j: i\in B_{l}; \atop j\in B_{r}; l<r}I(Y_{i,j}; Y_{j,i}).
\end{eqnarray*}
Therefore, the upper bound (\ref{upperbound2}) follows from
(\ref{upperbound}) and the above equality.
\end{proof}

The decomposition observed in the proof suggests that a graph based
approach can be used to study the pair-wise independent network. It
is our conjecture that the upper bound (\ref{upperbound2}) is always
tight for the pair-wise independent network; we demonstrate that
this conjecture holds in at least two special cases.

\section{The Broadcast Case}
In this section, we consider the broadcast case of a pair-wise
independent network in which the observations of each terminal in
$\{2,\cdots, m\}$ are correlated only with the observations of
terminal 1 (called the central terminal). In other words, the
observation $X_i$ by terminal $i\neq 1$ is equal to $Y_{i,1}$, and
$Y_{i,j}$ is a constant for $j\neq 1$.

By restricting ${\cal B}_k(A)$ to the set of 2-partitions
\[
(\{1, 3, \cdots, m\}, \{2\}), \cdots, (\{1, 2, \cdots, m-1\}, \{m\})
\]
in (\ref{upperbound2}), we obtain an upper bound on the secret key
capacity for the broadcast case
\begin{equation}
C_{SK}(\{1\cdots, m\})\leq \min_{2\leq i\leq m}I(Y_{1,i};Y_{i,1}).
\label{upperbound3}
\end{equation}
Next, we propose a protocol for the secret key establishment among
all $m$ terminals.

Terminals $2,\cdots, m$ begin by separately establishing secret keys
with the central terminal using the standard techniques
\cite{Mau93}, \cite{AhlCsi93}. This results in $m-1$ pair-wise
secret keys $K_{1,i}$, $2\leq i\leq m$, where $K_{1,i}$ denotes the
secret key shared by terminals 1 and $i$. Without loss of
generality, these keys are stored using a binary alphabet. Let
$|K_{1,i}|$ denote the length of the secret key $K_{1,i}$. According
to \cite{Mau93}, \cite{AhlCsi93}, for any $\epsilon >0$, each secret
key $K_{1,i}$, as a function of $(Y_{1,i}^n, Y_{i,1}^n)$, satisfies
the secrecy condition
\begin{equation}
I(K_{1,i};V_{1,i})\leq \epsilon,\label{secret}
\end{equation}
and the uniformity condition
\begin{equation}
H(K_{1,i})\geq |K_{1,i}|- \epsilon,\label{uniform}
\end{equation}
where $V_{1,i}$ denotes the public transmissions between terminal
$i$ and the central terminal to generate the pair-wise secret key
$K_{1,i}$. It follows from the independence condition
(\ref{independent}) that
\begin{equation}
I(K_{1,i};\{K_{1,j}: j\neq i\})\leq \epsilon.\label{independent2}
\end{equation}
The entropy rate of $K_{1,i}$ is given by \cite{Mau93},
\cite{AhlCsi93}
\begin{equation}
\frac{1}{n}H(K_{1,i})\geq I(Y_{1,i};Y_{i,1})-\epsilon.\label{rate}
\end{equation}

Let $K_{1,i^*}$, $2\leq i^*\leq m$, be the shortest key among the
$m-1$ generated keys, i.e., $|K_{1,i^*}|=\min_{2\leq i\leq m}
|K_{1,i}|$. This implies that
\begin{equation}
I(Y_{1,i^*};Y_{i^*,1})=\min_{2\leq i\leq m} I(Y_{1,i};Y_{i,1}).
\label{additional}
\end{equation}
The central terminal sends $\bar{K}_{1,i}\oplus K_{1,i^*}$ to
terminal $i$, where $\bar{K}_{1,i}$ denotes the first $|K_{1,i^*}|$
bits of $K_{1,i}$. At this point, all $m$ terminals have
$K_{1,i^*}$, which is set as the group secret key. The independence
between $K_{1,i^*}$ and all the public transmissions is shown in the
following proposition.

\begin{proposition}
For any $\delta>0$, the secret key $K_{1,i^*}$ generated above
satisfies
\begin{equation}
I(K_{1,i^*};\{V_{1,i}, K_{1,i^*}\oplus \bar{K}_{1,i} :2\leq i\leq
m\})\leq \delta. \label{lem1.0}
\end{equation}
\end{proposition}

\begin{proof}
In the interest of simple notation, we denote $K_{1,i^*}\oplus
\bar{K}_{1,i}$ by $\bar{V}_{1,i}$. Then the left side of
(\ref{lem1.0}) is written as
\begin{eqnarray}
&&I(K_{1,i^*};V_{1,2}, \cdots, V_{1,m}, \bar{V}_{1,2}, \cdots , \bar{V}_{1,m})\nonumber\\
&\leq &I(K_{1,i^*};\bar{V}_{1,2}, \cdots, \bar{V}_{1,m})\nonumber \\
&&+ I(K_{1,i^*},\bar{V}_{1,2},\cdots ,\bar{V}_{1,m}; V_{1,2},\cdots
,V_{1,m}). \label{lem1.2}
\end{eqnarray}
The former term in (\ref{lem1.2}) is upper bounded by
\begin{eqnarray}
\hspace{-0.15in} &&\hspace{-0.1in} I(K_{1,i^*};\bar{V}_{1,2}, \cdots, \bar{V}_{1,m})\nonumber\\
\hspace{-0.15in} &\leq& \hspace{-0.1in} \sum_{i=2\atop i\neq
i^*}^m\left[ H(\bar{K}_{1,i}\oplus K_{1,i^*})
-H(\bar{K}_{1,i}|K_{1,i^*},\bar{K}_{1,2},\cdots,\bar{K}_{1,i-1})\right]\nonumber\\
\hspace{-0.15in} &\leq & \hspace{-0.1in} 2(m-2)\epsilon, \nonumber
\end{eqnarray}
where the latter inequality follows from (\ref{uniform}) and
(\ref{independent2}). The latter term in (\ref{lem1.2}) is upper
bounded by
\begin{eqnarray}
&& I(K_{1,i^*},\bar{V}_{1,2},\cdots ,\bar{V}_{1,m}; V_{1,2},\cdots
,V_{1,m}) \nonumber\\
&=& I(K_{1,2},\cdots ,K_{1,m}; V_{1,2},\cdots ,
V_{1,m})\nonumber\\
&=& \sum_{i=2}^m I(K_{1,i};V_{1,i})\leq (m-1) \epsilon,\nonumber
\end{eqnarray}
where the inequality follows from (\ref{secret}). This completes the
proof.
\end{proof}

It follows from (\ref{rate}) and (\ref{additional}) that the
generated secret key $K_{1,i^*}$ has a rate close to the upper bound
(\ref{upperbound3}). Hence, the protocol is optimal. Furthermore, it
is not difficult to show that the protocol is also optimal for the
broadcast case with rate constraints (cf., \cite{CsiNar00}) on the
public transmissions.

\section{The Sub-group Key Case}

We now consider a sub-group key generation problem. Suppose that, in
a pair-wise independent network, terminals 1 and $m$ wish to
generate a secret key with the help of other $m-2$ terminals. In
other words, the sub-group $A=\{1,m\}$ of terminals wish to generate
a secret key.

We begin this section with a short overview of some definitions and
algorithms related to graphs. Then we propose a protocol for the
sub-group key generation problem. This protocol is based on existing
graph algorithms. Further, we show that the resulting secret key has
a rate close to the capacity.

Let ${\mathcal G}=({\mathcal N},{\mathcal E})$ be a weighted
directed graph. Let $s\in {\mathcal N}$ be a source node and $t\in
{\mathcal N}$ be a destination node in ${\mathcal G}$. An {\it $s-t$
cut} of the graph ${\mathcal G}$ is a partition of the nodes
${\mathcal N}$ into two sets ${\mathcal N}_1$ and ${\mathcal N}_2$
such that $s\in {\mathcal N}_1$ and $t\in {\mathcal N}_2$. Any edge
crossing from ${\mathcal N}_1$ to ${\mathcal N}_2$ is said to be a
{\it cut edge}. The {\it weight} of an $s-t$ cut is the sum of the
weights of its edges. An $s-t$ cut is {\it minimal} if the weight of
the $s-t$ cut is not larger than the weight of any other $s-t$ cut.

A {\it network flow} is an assignment of flow to the edges of a
weighted directed graph such that the amount of flow along the edge
does not exceed its weight. The maximal $s-t$ flow problem is to
find a maximal feasible flow from the source node $s$ to the
destination node $t$. The labeling algorithm \cite{Str86} is known
to solve the maximal $s-t$ flow problem.

By the max-flow min-cut theorem \cite{Sch03}, the maximal $s-t$ flow
is equal to the weight of the minimal $s-t$ cut.




We now return to the sub-group secret key generation problem. It
follows from Lemma 1 that the secret key capacity, which can be
achieved by terminals 1 and $m$ with the help of other terminals, is
upper bounded by
\begin{equation}
C_{SK}(\{1,m\})\leq \min_{(B_1,B_2)\in {\cal
B}_2(\{1,m\})}\sum_{i,j: i\in B_1; j\in B_2}I(Y_{i,j};
Y_{j,i}),\label{upperbound4}
\end{equation}
where ${\cal B}_2(\{1,m\})$ is the set of all 2-partitions of the
set $\{1,\cdots, m\}$ such that either atom of a 2-partition
intersects with $\{1, m\}$.

The upper bound (\ref{upperbound4}) can be represented via graphs.
Consider a weighted directed graph $G_1$ with $m$ nodes, each node
corresponding to a terminal. The edge from node $i$ to $j$ has
weight $I(Y_{i,j};Y_{j,i})$. Let node 1 be the source node and node
$m$ be the destination node. Then the upper bound
(\ref{upperbound4}) is equivalent to the minimal $s-t$ cut of $G_1$.

Next, we propose a protocol for the secret key establishment between
terminals 1 and $m$.

All the terminals begin by establishing pair-wise secret keys using
the standard techniques \cite{Mau93}, \cite{AhlCsi93}. This results
in $\left(m\atop 2\right)$ pair-wise secret keys. Let $K_{i,j}$
($=K_{j,i})$ denote the secret key shared by terminals $i$ and $j$.
Each secret key $K_{i,j}$, as a function of $(Y_{i,j}^n,
Y_{j,i}^n)$, satisfies certain secrecy condition and uniformity
condition as in (\ref{secret}), (\ref{uniform}). Further, for any
$\epsilon>0$,
\begin{equation} I(K_{i,j};\{K_{k,l}:
(k,l)\neq (i,j), (j,i)\})\leq \epsilon,\label{independent3}
\end{equation}
and the entropy rate of $K_{i,j}$ is given by \cite{Mau93},
\cite{AhlCsi93}
\begin{equation}
\frac{1}{n}H(K_{i,j})\geq I(Y_{i,j};Y_{j,i})-\epsilon. \label{key}
\end{equation}
Based on the pair-wise secret key $K_{i,j}$, terminal $i$ can cipher
$|K_{i,j}|$ random bits with $K_{i,j}$ through the one-time pad
before transmitting these random bits to terminal $j$ (and vice
versa). This implies the existence of a secure channel between nodes
$i$ and $j$ with capacity $\frac{1}{n}|K_{i,j}|$.

Consider a weighted directed graph $G_2$ with $m$ nodes, each node
corresponding to a terminal. The weight of an edge $(i,j)$ in the
graph is equal to the capacity of the secure channel connecting
terminals $i$ and $j$, i.e., $\frac{1}{n}|K_{i,j}|$. Using the
labeling algorithm \cite{Str86}, one can find the maximal $s-t$ flow
$F$ in this graph. Accordingly, terminal 1 can securely send random
bits through the network to terminal $m$ at rate $F$. Let these
random bits be the secret key of terminals 1 and $m$. By arguments
similar to those used in the proof of Proposition 1, it is easy to
show that this secret key is nearly statistically independent of the
public transmissions.
\begin{proposition}
Let $V$ denote all the public transmissions needed in the protocol
above. For any $\delta>0$, the secret key $K$ generated above
satisfies $I(K;V)\leq \delta$.
\end{proposition}
\vspace{0.1in}

According to the max-flow min-cut theorem \cite{Sch03}, the rate $F$
of the generated secret key is equal to the minimal $s-t$ cut of
$G_2$. It follows from (\ref{key}) that the minimal $s-t$ cut of
$G_2$ is close to the minimal $s-t$ cut of $G_1$. Hence, the
achieved secret key rate is close to the upper bound
(\ref{upperbound4}), and the protocol is optimal.

\section{The Group Key Case}
In this section, we examine the problem of all the terminals in a
pair-wise independent network generating a common secret key. We
start by a short overview of more definitions and algorithms related
to graphs. Then we propose a protocol for the group secret key
generation problem. This protocol is based on existing graph
algorithms. Finally, we demonstrate the efficiency of this protocol
through several examples.

Let ${\mathcal G}=({\mathcal N},{\mathcal E})$ be a weighted
undirected graph. The graph ${\mathcal G}$ is said to be {\it
connected} if for every two distinct nodes $i, j\in {\mathcal N}$,
there exists a path from node $i$ to node $j$. Otherwise, the graph
is said to be {\it unconnected}. Define a {\it multi-cut} of
${\mathcal G}$ to be a partition of the nodes ${\mathcal N}$ into
several sets ${\mathcal N}_1,\cdots ,{\mathcal N}_L$, $2\leq L\leq
m$, with $m$ being the number of nodes in ${\mathcal G}$. Any edge
$(i,j)\in {\mathcal E}$ with end-nodes $i$, $j$ belonging to
different sets is said to be a {\it multi-cut edge}. The {\it
weight} of a multi-cut is the weight sum of its edges. The {\it
normalized weight} of a multi-cut is the weight of the multi-cut
divided by $L-1$, where $L$ is the number of sets in the partition
of ${\mathcal G}$ generating the multi-cut.

Given a connected undirected graph ${\mathcal G}=({\mathcal
N},{\mathcal E})$, let ${\mathcal E}_1$ be a subset of ${\mathcal
E}$ such that ${\mathcal T}=({\mathcal N},{\mathcal E}_1)$ is a
tree. Such a tree is called a {\it spanning tree}. A {\it maximum
spanning tree} from a weighted graph is defined as a spanning tree
such that the weight sum of its edges is as large as possible. The
problem of finding a maximum spanning tree can be solved by several
greedy algorithms. Two examples are Kruskal's algorithm and Prim's
algorithm (cf., e.g., \cite{Sch03}).

The upper bound (\ref{upperbound2}) on the secret key capacity for
the group secret key case, i.e., $A=\{1,\cdots ,m\}$, can be
represented via graphs. Consider a weighted undirected graph $G_3$
with $m$ nodes, each node corresponding to a terminal. The weight of
an edge $(i,j)$ in the graph is equal to $I(Y_{i,j};Y_{j,i})$. Note
that each multi-cut of the graph $G_3$ is equivalent to a partition
in (\ref{upperbound2}), and the set of all multi-cuts of the graph
$G_3$ is precisely equivalent to the set of partitions $\{(B_1,
\cdots ,B_k)\in {\cal B}_k(\{1, \cdots ,m\}): 2\leq k\leq m\}$ in
(\ref{upperbound2}). Moreover, the normalized weight of a multi-cut
is precisely $\frac{1}{k-1} I_k'(\{1, \cdots, m\})$. Consequently,
we have the following corollary.
\begin{corollary}
The secret key capacity for the group secret key case is upper
bounded by the minimal normalized weight of the multi-cuts of $G_3$.
In particular, this upper bound implies the following two upper
bounds:

i). the minimal weight of the cuts of $G_3$, where a cut is a
multi-cut generated by a partition into 2 sets;

ii). the weight sum of all edges in $G_3$ divided by $m-1$.
\end{corollary}

\vspace{0.1in}

Next, we propose a protocol for the group secret key generation
problem. All the terminals begin by establishing pair-wise secret
keys using the standard techniques \cite{Mau93}, \cite{AhlCsi93}.
Let $K_{i,j}$ ($=K_{j,i}$) denote the secret key shared by terminals
$i$ and $j$. These secret keys satisfy the certain secrecy
condition, uniformity condition, and (\ref{independent3}),
(\ref{key}).

Consider a weighted undirected graph $G_4$ with $m$ nodes, each
corresponding to a terminal. The weight of an edge $(i,j)$ in the
graph is equal to the length\footnote{For the purpose of simple
notations, we shall use the length, rather than the rate, of a
secret key as an edge weight. This should not lead to any
confusion.} of the corresponding pair-wise secret key $K_{i,j}$,
i.e., $|K_{i,j}|$.

Our group key generation algorithm is related to Lemma 2 below. This
lemma discusses the generation of a single secret bit among $m$
nodes, based on a single bit from each of the $m-1$ pair-wise secret
keys whose corresponding edges constitute a spanning tree.

\begin{lemma}
Consider an arbitrary tree connecting $m$ nodes. If every pair of
neighbor nodes on the tree shares a single pair-wise secret bit,
then a single secret bit can be generated among all $m$ nodes.
\end{lemma}

\begin{proof}
A simple algorithm on generating a single secret bit among all $m$
nodes is illustrated below.

\noindent {\bf Single Bit Algorithm}:

{\bf Step 1}. Randomly pick up an edge $(i^*,j^*)$ from the spanning
tree. Nodes $i^*$ and $j^*$ share a secret bit $B_{i^*,j^*}$.

{\bf Step 2}. If node $i$ knows $B_{i^*,j^*}$, but its neighbor node
$j$ does not, then node $i$ sends $B_{i^*,j^*}\oplus B_{i,j}$ to
node $j$, where $B_{i,j}$ is the secret bit shared by nodes $i$ and
$j$. Upon receiving this message, node $j$ is able to decode
$B_{i^*,j^*}$. Repeat this step until the above condition does not
hold. $\hfill\Box$

This algorithm stops when all the nodes are able to decode
$B_{i^*,j^*}$. It is trivial to show the independence between
$B_{i^*,j^*}$ and the public transmissions. Hence, $B_{i^*,j^*}$ is
a secret bit.
\end{proof}

Our group secret key generation algorithm is given below.

\noindent {\bf Group Key Generation Algorithm}:

Let $G$ be the weighted undirected graph $G_4$ defined above.

{\bf Step 1}: Determine a maximum spanning tree from $G$, using any
known algorithm (e.g., Kruskal's or Prim's). If there is more than
one maximum spanning tree, randomly select one.

{\bf Step 2}: Apply the single bit algorithm to generate a single
secret bit among all nodes, based on a single bit from every
pair-wise secret key on the determined maximum spanning tree. Note
that these used bits will be of no use in the remaining group key
generation process.

{\bf Step 3}: Update the graph by reducing the edge weight by 1 for
the edges on the determined spanning tree. Remove an edge when its
weight becomes zero.

{\bf Step 4}: If the remaining graph $G$ is unconnected, then set
the group secret key as the collection of all generated secret bits.
Otherwise, return to Step 1. $\hfill\Box$

Since each iteration of the group key generation algorithm leads to
a single secret bit, the length of the resulting secret key is equal
to the number of iterations of the algorithm that can be run until
the graph becomes unconnected. The purpose of searching a maximum
spanning tree (rather than picking up an arbitrary spanning tree) in
Step 1 is to maximize the number of iterations of the algorithm by
means of ``balancing'' edge weights in the weight reduction
procedure.

By arguments similar to those used in the proof of Proposition 1, it
is easy to show that the secret key resulting from the above
algorithm is nearly statistically independent of the public
transmissions.
\begin{proposition}
Let $V$ denote all the public transmissions needed in the protocol
above. For any $\delta>0$, the secret key $K$ generated above
satisfies $I(K;V)\leq \delta$.
\end{proposition}

\vspace{0.1in}

We illustrate the operations of the group key generation algorithm
through the following example.

\begin{example}
Consider a network with 3 nodes. Nodes 1 and 2 share a secret key of
5 bits; nodes 1 and 3 share a secret key of 4 bits; and nodes 2 and
3 share a secret key of 3 bits. This network is drawn in the left
part of Fig. 1.

\begin{figure}
\hspace{-0.2in} \scalebox{0.349}{\includegraphics{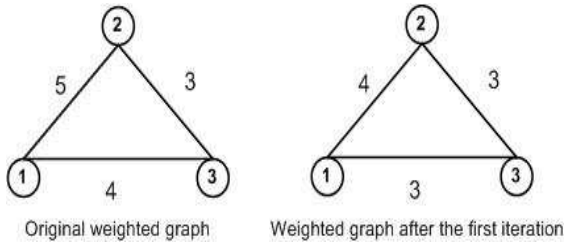}}
\caption{Example network with 3 nodes} \label{Fig1}
\end{figure}

Let the pair-wise secret keys be $K_{1,2}=(K_{1,2}^1, \cdots,
K_{1,2}^5)$, $K_{1,3}=(K_{1,3}^1, \cdots, K_{1,3}^4)$, and
$K_{2,3}=(K_{2,3}^1, \cdots, K_{2,3}^3)$, where $K_{i,j}^k$ denotes
the $k^{th}$ bit of the secret key shared by nodes $i$ and $j$.

The spanning tree $((1,2),\ (1,3))$ is the maximum spanning tree
from the graph in the left part of Fig. 1, as it has a larger weight
(= 9) than other spanning trees. Hence, by the single bit algorithm,
node 1 transmits $K_{1,2}^1\oplus K_{1,3}^1$ and sets $K_{1,2}^1$
(or $K_{1,3}^1$) as the secret bit. Update the graph by reducing the
weights of the edges $(1,2)$, $(1,3)$ by 1. This results in the
graph given in the right part of Fig. 1.


By repeating the above process, the determined maximum spanning
trees and the corresponding public transmissions in the next five
iterations are
\[((1,2),\ (1,3)),\ \ \  ((1,2),\ (2,3)),\ \ \  ((1,2),\ (2,3)),\]
\[((1,3),\ (2,3)),\ \ \  ((1,2),\ (1,3)),\]
and
\[K_{1,2}^2\oplus K_{1,3}^2,\ \ \ K_{1,2}^3\oplus K_{2,3}^1,
\ \ \ K_{1,2}^4\oplus K_{2,3}^2,\]
\[K_{1,3}^3\oplus K_{2,3}^3,\ \ \ K_{1,2}^5\oplus K_{1,3}^4,\]
respectively. The algorithm stops after these iterations, as the
remaining graph is unconnected. The group secret key is set as $
(K_{1,2}^1,K_{1,2}^2, K_{1,2}^3, K_{1,2}^4, K_{1,3}^3, K_{1,2}^5)$.
By restricting $k$ to $|A|=3$ and setting $Y_{i,j}=Y_{j,i}=K_{i,j}$
in (\ref{upperbound2}), we find that the length of any group secret
key in this example cannot be larger than 6 bits. Hence, the
algorithm is optimal.

For a network with 3 nodes, determining a maximum spanning tree in
the group key generation algorithm is equivalent to determining a
node such that the weight sum of two edges connecting with this node
is the largest.
\end{example}



\begin{example}
Consider a network with $m$ nodes and all $\left( m\atop 2 \right)$
edges having the same even weight $w=2u$, for a certain positive
integer $u$. A secret key of length $mu$ bits can be generated by
using the group key generation algorithm. On the other hand, by
restricting $k$ to $|A|=m$ and setting $Y_{i,j}=Y_{j,i}=K_{i,j}$ in
(\ref{upperbound2}), we find that the length of any group secret key
in this example cannot be larger than $\frac{w\left( m\atop 2
\right) }{m-1}=mu$ bits. Hence, the algorithm is optimal.
\end{example}

Although the group key generation algorithm is shown to be optimal
in the examples above, its potential non-optimality is demonstrated
by the following example.

\begin{example}
Consider a network with 4 nodes. Each node is connected with every
other node by an edge of weight 1. It is clear that $((1,2),\
(1,3),\ (1,4))$ is a maximum spanning tree of the graph, which means
that 1 secret bit can be generated from it. However, the updated
graph then becomes unconnected, resulting in a secret key of 1 bit.

Nevertheless, the upper bound in (\ref{upperbound2}) can be achieved
by simply making a \emph{better selection from the possible maximal
spanning trees}. One such tree is $((1,2),\ (2,3),\ (3,4))$. After
the weight reduction, the new graph is still connected, having the
spanning tree $((1,3),\ (1,4),\ (2,4))$. Hence, 2 secret bits, which
is optimal, can be established in this manner.
\end{example}

This example suggests the importance of deliberately selecting a
maximum spanning tree in Step 1 of the algorithm. What a good
selection scheme might look like, and whether it would guarantee the
optimality of this algorithm, remains open.

\vspace{-0.05in}
\section*{Acknowledgment}
The authors would like to thank Yogendra Shah and Inhyok Cha for
introducing the sub-group key generation problem and for helpful
discussions on this topic.

\vspace{-0.02in}




%

\end{document}